\documentclass[aps,prd,reprint,longbibliography,notitlepage,superscriptaddress,preprintnumbers,floatfix,nofootinbib]{revtex4-2}
\usepackage[T1,T2A]{fontenc}
\usepackage[utf8]{inputenc} 
\usepackage{mathrsfs}
\usepackage{amssymb,amsmath,mathtools}
\usepackage{amsthm}

\usepackage{graphicx}
\usepackage{xcolor}
\usepackage{placeins}
\usepackage[normalem]{ulem}
\usepackage{stmaryrd}
\usepackage{fmtcount}

\definecolor{lcolor}{rgb}{0.5,0,0}
\definecolor{citcolor}{rgb}{0,0.3,0.0}
\definecolor{coloryksi}{rgb}{0.5,0.0,0.0}
\definecolor{colorkaksi}{rgb}{0.0,0.0,0.5}
\definecolor{colorkolme}{rgb}{0.0,0.3,0.1}

\usepackage{physics}
\usepackage{tensor}
\usepackage{cancel}
\usepackage{mathrsfs}

\newcommand{\beq}{\begin{equation}}
\newcommand{\eeq}{\end{equation}}

\newcommand{\xbj}{{x_\mathrm{Bj}}}

\DeclareMathOperator{\argmin}{arg\,min}

\definecolor{jyuorange}{RGB}{241,86,63}

\newcommand{\edit}[1]{{\color{black}#1}}

\usepackage[breaklinks,colorlinks,urlcolor=blue,citecolor=citcolor,linkcolor=lcolor]{hyperref}

\newcommand{\rt}{{\mathbf{r}}}

\newcommand{\bt}{{\mathbf{b}}}

\newcommand{\besk}{{\mathrm{K}}}

\newcommand{\ud}{\, \mathrm{d}}

\newcommand{\nc}{{N_\mathrm{c}}}

\newcommand{\gev}{\ \textrm{GeV}}

\newcommand{\aem}{\alpha_{\mathrm{em}}}

\newcounter{diag}
\newcounter{subdiag}[diag]
\begin{document}
\allowdisplaybreaks

\author{H. H\"anninen}
\affiliation{
Department of Mathematics and Statistics, University of Jyväskylä,
 P.O. Box 35, 40014 University of Jyv\"askyl\"a, Finland}

\author{A. Kykk\"anen}
 \affiliation{Department of Computational Applied Mathematics and Operations Research,
 \\
 Rice University, Houston TX, USA}

\author{H. Schl\"uter}
\affiliation{
Department of Mathematics and Statistics, University of Helsinki,
P.O. Box 3, 00014 University of Helsinki, Finland}

\title{Reconstruction of the Dipole Amplitude in the Dipole Picture as a mathematical Inverse Problem}

\begin{abstract}
We show that the inference problem of constraining the dipole amplitude with inclusive deep inelastic scattering data can be written into a discrete linear inverse problem, in an analogous manner as can be done for computed tomography.
To this formulation of the problem, we apply standard inverse problems methods and algorithms to reconstruct known dipole amplitudes from simulated reduced cross section data with realistic precision.
The main difference of this approach to previous works is that this implementation does not require any fit parametrization of the dipole amplitude.
The freedom from parametrization also enables us for the first time to quantify the uncertainties of the inferred dipole amplitude in a novel more general framework.
This mathematical approach to small-$x$ phenomenology opens a path to parametrization bias free inference of the dipole amplitude from HERA and Electron--Ion Collider data.
\end{abstract}

\maketitle

\section{Introduction}
    
    Quantum chromodynamics (QCD) is the part of the Standard Model that describes the strong nuclear force and its elementary particles quarks and gluons.
    A remarkable fact about the development of QCD is that every model building step during its inception---such as figuring out what hadrons are made of---had to be taken in the virtual darkness of the directly unobservable.
    In the 1950s advancements in particle detectors lead to experimental discovery of numerous particles called hadrons~\cite{ParticleDataGroup:2024cfk}, in response to which began the development of theories which would explain the observed particles as composed of smaller particles called quarks~\cite{Gell-Mann:1961omu,Gell-Mann:1964ewy,Zweig:1964ruk,Zweig:1964jf,Neeman:1961jhl,Bjorken:1964gz} or partons~\cite{Feynman:1969ej}.
    The first experimental evidence for smaller elementary particles within the proton would only emerge later as the inference of the existence of partons at SLAC~\cite{Bloom:1969kc,Breidenbach:1969kd}, and which later were confirmed to agree with the parton model predictions.
    
    In the decades since the first formulation of QCD, the high-energy physics program has deepened our understanding of the interior structure of the proton, hadrons in general, and the precision quantification of the highly non-trivial strong nuclear force~\cite{Gross:2022hyw}. Deep theoretical understanding of many phenomena still remains elusive, such as color confinement~\cite{Greensite:2011zz}, and gluon saturation expected at high energy scales \cite{Morreale:2021pnn}.
    The gluon density within the proton is understood to grow significantly at high energy. It is also theoretically expected that the self-interactions of the gluons will at some energy scale begin tempering this growth, leading to a maximal density---this is the phenomenon known as saturation in high energy QCD.
    Precision theory understanding of gluon saturation is one of the central objectives of the Electron--Ion Collider~\cite{Accardi:2012qut,AbdulKhalek:2021gbh,Abir:2023fpo} in construction, and would be a key opportunity with the proposed future accelerator facilities LHeC~\cite{LHeC:2020van} and EicC~\cite{Anderle:2021wcy}.
    These particle accelerator experiments collide high-energy electrons with protons and heavy-ions in deep inelastic scattering (DIS), which enables the experiment to be sensitive to the internal structure of the target proton or heavy-ion. The electron is a point-like elementary particle and so it is a nice clean probe to study the rich hadronic structure of the targets.
    With this opportunity in mind, we apply the inverse problems paradigm to develop a more general methodology for the inference of the features of saturation from deep inelastic scattering data.
    Specifically, we work in the dipole picture of DIS, which is a theory description of electron--proton scattering valid at extremely high energies, and develop mathematical methodology for the inference of a non-perturbative quantity called the dipole amplitude.

    Inverse problems is a field of mathematics, which was developed to enable indirect measurement in scientific and engineering applications~\cite{Uhlmann2003InsideO}.
    One of the first applications of the inverse problems perspective is considered to be Le Verrier's prediction for the existence of the planet Neptune in the 19th century from the orbital anomalies Uranus~\cite{neptune:lassell:1846,neptune:airy:1846,neptune:adams:1846}.
    Since then the field of inverse problems has grown to encompass numerous applications in physical and applied sciences~\cite{Uhlmann2003InsideO,Argoul:2012inverse:overview}---non-exhaustively we mention: acoustics, calorimetry, geophysics, imaging, meteorology, (non-)destructive testing, oceanography, optics, radar, radioastronomy, spectroscopy.
    Contrast these applications with the areas of the development of the mathematical theory of inverse problems~\cite{Uhlmann2003InsideO,Hansen:2021ct,Salo:2023inverse:pde,salo2019applicationsmicrolocalanalysisinverse}: integral transformation inverse problems, such as the Radon transform, inverse problems involving partial differential equations such as the wave equation, the conductivity equation or the heat equation, and geometric inverse problems on manifolds.
    We highlight this aspect of the field of inverse problems to pin down a subtle feature of the nature of inverse problems: inverse problems study the mathematical connection between a measurement and a directly unobservable---inferrable---quantity.
    Critically, the nature of this inferrable--measurement connection is not inherently tied to the physics of the process, as we can perhaps infer from the varied applications and the similarities between the inverse problems in their broader mathematical categories of inverse problems.
    
    Consider as an example the inverse problem of computed tomography~\cite{Hansen:2021ct,buzug2009computed,deans1983radon,kak2001:principlesct}.
    The relevant physical process is the propagation and attenuation of x-rays a medium, which is measured in a sinogram that records the projections of the target structure from multiple directions. 
    The relevant mathematical inverse problem is the reconstruction of the interior structure of the target from the sinogram, which is a linear integral transformation inverse problem, and the solution of which is not tied to the physical process, but to the mathematical nature of the connection between the internal structure and the measurement.
    One of our foundational observations in this work is that the mathematical understanding of this inferrable--measurement connection can be applied to phenomenology in QCD to enable inference of non-perturbative quantities in a novel manner and with reduced or eliminated parametrization bias.

    We show in this article that the inference problem for the dipole amplitude from DIS data can be formulated as a linear integral transformation inverse problem, which is a large category of inverse problems with robust and general solution algorithms \cite{Hansen:2017airtools,Hansen:2021ct}. An example of a well understood inverse problem from that class is computed tomography.
    We apply this inverse problems approach to QCD phenomenology by writing the inference problem into an explicit reconstruction problem. For this explicit problem, we then implement a numerical closure test where we generate reduced cross section data for DIS and show that standard methods for this inverse problem type are able to reconstruct the known fit dipole amplitudes from the generated data without any fit parametrization.
    This freedom from parametrization bias enables us for the first time evaluate the uncertainty of the inference of the dipole amplitude in a more general manner than has been possible previously, and the regimes of large uncertainties match the physical theory based expectations. Recent work~\cite{Casuga:2023dcf,Casuga:2025etc} improves the robustness of the uncertainty quantification beyond previous approaches---cf. Refs~\cite{Beuf:2020dxl,Hanninen:2021byo,Hanninen:2022gje} and the references therein---by using Bayesian analysis of the inference based on a theoretically motivated fit parametrization of the dipole amplitude, whereas in this work we write the inference problem into a parametrization-free inverse problem.
    This freedom from parametrization---and the uncertainty quantification enabled by it---is in principle comparable to the approach employed by the NNPDF collaboration, who use neural networks to infer the parton distribution functions from particle collision measurements~\cite{NNPDF:2020yuc,NNPDF:2021whr,NNPDF:2022qks,NNPDF:2024dpb}.
    \edit{The present work also shares a similarity in approach with Ref.~\cite{Munier:2001nr}, where the impact parameter dependence of the dipole amplitude is solved from vector meson electroproduction data by leveraging the analytic invertibility of the Fourier transform, whereas in this work we apply general methods to invert the specific integral transform present in the inclusive cross section formulae to solve the dipole amplitude.}
    
    This work opens a straightforward path towards reconstruction of the dipole amplitude from HERA data without parametrization bias where uncertainties to the reconstruction are propagated both from the experimental data and perturbation theory. This reconstruction would be akin to an indirect measurement like it is the case for the reconstruction of the internal structure of the medium in computed tomography.
    This approach would enable novel quantified analysis of tension between established inferred parametrizations of the dipole amplitude and data---such as the large anomalous dimension values seen in Ref.~\cite{Beuf:2020dxl}---which has the potential to be a powerful methodology in the upcoming era of precision measurements enabled by the EIC.

    In the closure test, we use the dipole amplitude parametrizations determined in a Bayesian inference framework~\cite{Casuga:2023dcf} as the known unknown to generate the data, and then reconstruct these dipole amplitudes from the data.
    Conventionally, the dipole amplitude is constrained with data by fitting a parametrization to data, such as in Refs.~\cite{GolecBiernat:1998js,Bartels:2002cj,Kowalski:2003hm,Iancu:2003ge,Kowalski:2006hc,Watt:2007nr,Rezaeian:2012ji,Mantysaari:2018nng,Albacete:2009fh,Albacete:2010sy,Lappi:2013zma,Ducloue:2019jmy,Soyez:2007kg,Berger:2011ew,Rezaeian:2013tka,Bendova:2019psy,Beuf:2020dxl,Hanninen:2022gje,Casuga:2023dcf,Casuga:2025etc}, using phenomenologically derived parametrizations such as the McLerran--Venugopalan~\cite{McLerran:1993ni}, Golec-Biernat--W\"usthoff~\cite{GolecBiernat:1998js}, or IPsat~\cite{Kowalski:2003hm} model.
    In our formalism this inference-by-fit is an approach to solve the implicit inverse problem.
    The state-of-the-art solutions of the implicit inverse problem have improved the theory precision of the direct problem to next-to-leading order (NLO) accuracy, which requires extensive quantum field theoretic calculations of the deep inelastic scattering~\cite{Beuf:2016wdz, Beuf:2017bpd, Hanninen:2017ddy, Beuf:2021qqa, Beuf:2021srj, Beuf:2022ndu}, and similar progress---including theory calculations of next-to-eikonal effects---has been made with other related high-energy scattering processes as well~\cite{Iancu:2016vyg,Ducloue:2017mpb, Stasto:2013cha,Stasto:2014sea,Altinoluk:2014eka,Kang:2014lha,Altinoluk:2015vax,Boussarie:2016ogo,Boussarie:2016bkq,Mantysaari:2021ryb,Mantysaari:2022kdm,Beuf:2022kyp,Beuf:2024msh,Abir:2023fpo, Altinoluk:2022jkk, Altinoluk:2023qfr, Altinoluk:2024dba, Altinoluk:2024tyx, Altinoluk:2024zom, Hatta:2022lzj, Agostini:2024xqs, Fucilla:2022wcg, Fucilla:2023mkl}.
    The state-of-the-art of the inference of the dipole amplitude use the NLO accuracy theory description of the scattering to constrain the parametrization~\cite{Beuf:2020dxl,oma3:github,oma3:zenodo,Hanninen:2022gje,Casuga:2025etc}.
    The reduction of the parametrization bias enabled by the mathematical inverse problems approach of this work might lead to new insight into some challenges of the existing parametrizations, such as the Fourier transform positivity problem~\cite{Casuga:2023dcf,Casuga:2025etc}.

    Some questions related to the mathematical inverse problems framework related to this work are left open and for future work. One of such questions is, whether the inverse problem is well-posed. Well-posedness is classified in terms of the Hadamard criteria~\cite{hadamard}, i.e. that there exists a solution with any given measurement set, that the solution is unique, and that the solution continuously depends on the measured dataset. If any of these criteria are not met, the problem is said to be ill-posed. However, this is not a problem inherently or for the application in this work, since many applied inverse problems are ill-posed, such as the computed tomography example discussed, which fails both in the continuity and surjectivity criteria~\cite{Hansen:2021ct}. These aspects are considered in the discussions of the implementation and results in the relevant sections.
    Relatedly, while it would be preferable to use the NLO accuracy description of the scattering process for the forward problem, it complicates the formulation of the explicit inverse problem as the inverse problem becomes non-linear. Thus it is left for future work, and we first construct this approach using the leading order accuracy description of DIS in the dipole picture.

    This article is structured as follows: 
    In Sec.~\ref{sec:dis} we describe the well-known forward problem of virtual photon--proton DIS, and discuss how the standard inference approach is an implicit inverse problem.
    Then in Sec.~\ref{sec:inverse} we discuss how the problem can be written into an explicit integral transformation inverse problem to which we may apply standard solution methods, and in Sec.~\ref{sec:results} we discuss the numerical implementation of the solution and results. Finally, in Sec.~\ref{sec:conclusions}, we finish with a discussion of conclusions, future work, and potential impact of the underlying methodology in high-energy phenomenology.
    
    \textit{N.B.} Since the readership of this article is intended to include both high-energy physicists and inverse problems mathematicians, we strive to cover key discussions in a more careful and inclusive detail than might be typical.

\section{The forward problem of deep inelastic scattering in the dipole picture} \label{sec:dis}

    \begin{figure}[tb]
        \centering
        \includegraphics[width=\columnwidth]{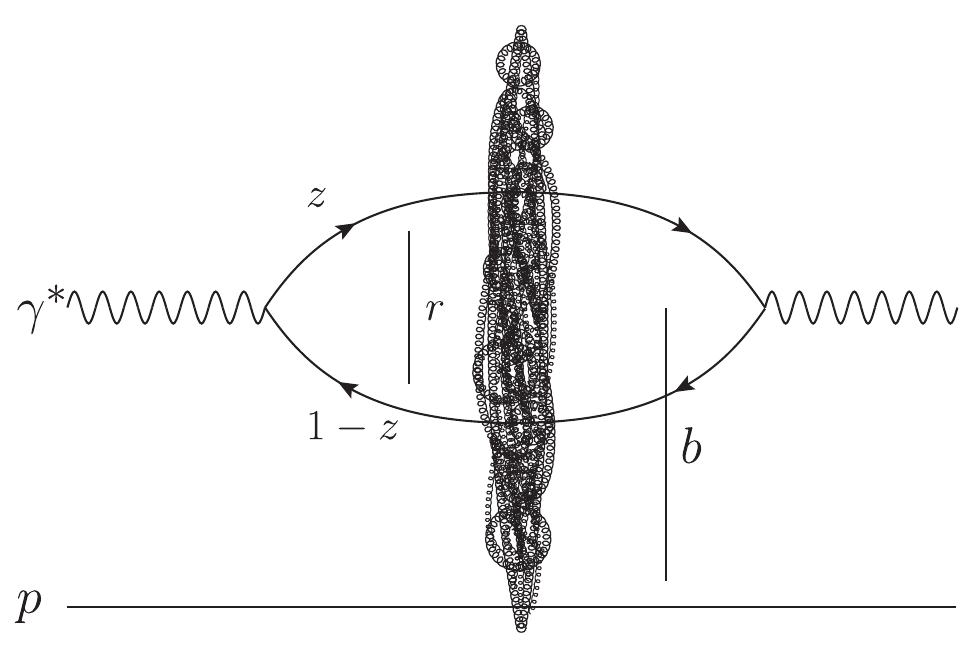}
        \caption{Depiction of the virtual photon--proton deep inelastic scattering in the dipole picture. At leading order in light-cone perturbation theory, the incoming virtual photon quantum state fluctuates into a quark--antiquark state, which is able to probe the strong nuclear force field of the proton.
        In this figure, time proceeds left to right, and vertical axis is the separation between the probe $\gamma^*$ and target proton, which are primarily traveling in opposite directions in the collider experiment. Key quantities are the transverse size of the quark--antiquark dipole-state $r$, the transverse separation of the target and the incoming quantum state $b$, and the fractional momentum of the quarks in the state $z$.
        The dipole amplitude---the quantity which describes the scattering process of the dipole-state off the target shown in the center---is to be inferred from the collider experiment data.}
        \label{fig:diag-lo-dis}
    \end{figure}

    We begin by describing the forward problem of electron--proton deep inelastic scattering at leading order in the dipole picture.
    A collision between the electron and proton is called deep inelastic scattering (DIS), when the interaction between the particles is strong enough to break the proton apart, which is then detected as showers of particles in the particle accelerator experiment. The fact that the proton dissociates implies that the scattering process becomes sensitive to the internal structure of the proton, which is in contrast with elastic scattering, where the electron and proton scatter intact, and are seen as such in the detector.
    In basic terms, particle collider experiments such as DESY-HERA and Electron--Ion Collider form and accelerate a beam of electrons and a beam of protons traveling in opposite directions, which are then collided within various detectors that track and record particles that escape from the collisions. The data is then analyzed to deduce for example the total probability of a collision at a given energy of the collision.
    Numerous kinds of inverse problems then arise with the task of inferring the internal structure of the proton, or precision understanding of the strong nuclear force via quantum chromodynamics.
    The dipole picture~\cite{Mueller:1993rr,Mueller:1994gb,Mueller:1994jq,Chen:1995pa} is a theory description of the electron--proton scattering valid at extremely high-energies, where it is known that the gluon content of the proton becomes so high that it overshadows the presence of the familiar valence quarks. Therefore the scattering of the electron---via an intermediate quantum state describing the interaction between the probe and target---is described in the dipole picture to happen off the gluon cloud within the proton.

    The experimentally observed total probability of scattering is reported in terms of a related quantity called \emph{reduced cross section} $\sigma_r$~\cite{Accardi:2012qut, Abramowicz:2015mha,H1:2018flt,Aaron:2009aa}, where the cross section is defined as the quantum mechanical analogue of the transverse cross sectional area of the target~\cite{Peskin:1995ev}.
    The reduced cross section is defined in terms of the proton structure functions $F$:
    \begin{align}
        \sigma_r(y,x,Q^2) & = F_2(x,Q^2) - \frac{y^2}{1+(1-y)^2}F_L(x,Q^2) 
        \notag \\ \label{eq:sigmared}
        & = F_T(x,Q^2) + \frac{2(1-y)}{1+(1-y)^2}F_L(x,Q^2)
    \end{align}
    where $x$ is the Bjorken-$x$, $Q^2$ is the virtuality of the incoming virtual photon interaction carrier emitted by the electron probe, and $y \coloneqq \frac{Q^2}{s x} \in (0,1)$ is the inelasticity of the scattering. The Mandelstam variable $s$ is related to the center-of-mass energy of the particle collision by $s=E_{CM}^2$. 
    The longitudinal and transverse structure functions of the proton, $F_L$ and $F_T$ respectively, are defined as
    \begin{equation} \label{eq:structure-functions}
    F_{L,T}(x,Q^2) = \frac{Q^2}{4 \pi \aem} \sigma_{L,T}(x,Q^2),
    \end{equation}
    where $\sigma_{L,T}(x,Q^2)$ are the total cross sections for a $L$ longitudinally and $T$ transversely polarized virtual photon--proton scattering, and $\aem$ is the fine structure constant---the quantity describing the strength of electromagnetic interaction in quantum electrodynamics. Another important structure function is $F_2 \coloneqq F_T + F_L$, which is proportional to the total (unpolarized) cross section $\sigma_\mathrm{tot} \coloneqq \sigma_T + \sigma_L$.

    Quantum chromodynamics---specifically an effective theory in the high-energy limit called color glass condensate effective field theory~\cite{Kogut:1969xa, Bjorken:1970ah, Lepage:1980fj, Brodsky:1997de, Kovchegov:2012mbw, Iancu:2003xm, Weigert:2005us, Gelis:2010nm, Albacete:2014fwa, Blaizot:2016qgz, Morreale:2021pnn} for this work---enables the calculation of the DIS total scattering cross sections $\sigma_{L,T}(x,Q^2)$ based on the knowledge of the in- and out-going elementary particles~\cite{Kovchegov:2012mbw}.
    This theory picture of the scattering is illustrated in Figure~\ref{fig:diag-lo-dis}, which to a quantum field theorist depicts the elements that go into the calculation of the total cross section $\sigma_{L,T}$.
    The resulting $T$ and $L$ cross sections at leading order (LO) accuracy are
    \begin{multline}
        \label{eq:dis-cross-section}
        \sigma_{T,L}^{\gamma^*p \to X} \!(x, Q^2) =
            \frac{1}{4\pi} \sum_f
            \int_{\mathbb{R}^2} \int_0^1
            \abs{\psi_{T,L}^{\gamma^* \to q_f \bar q_f}(\rt, Q^2, z, f)}^2
            \\
            \times
            \sigma_{q \bar q}(r,x)
            \ud^2 \rt \ud z,
    \end{multline}
    where $\rt$ is the transverse separation of the $q \bar q$-dipole, $z$ is the longitudinal momentum fraction of the quark $q_f$, and $f$ is the flavor of the quark dipole scattering off the target.

    To elaborate on the parts that go into Eq.~\eqref{eq:dis-cross-section}, let us on a high-level discuss where they come from, and how this allows us to reformulate the right hand side of the equation into a forward operator that acts on the unknown quantity to be inferred as the solution of the inverse problem.

    First, let us consider the $|\psi|^2$-term acting as the kernel in the integral operation in Eq.\eqref{eq:dis-cross-section}.
    In the dipole picture of lepton--proton scattering, the lepton scatters by emitting a virtual photon $\gamma^*$, which then interacts with the target proton. Since the virtual photon does not carry color charge, it first fluctuates into a quantum state which does---which in the leading contribution to the quantum field theoretic calculation is a quark--antiquark dipole---and that state can then scatter off the strong nuclear force field of the proton, illustrated in Fig.~\ref{fig:diag-lo-dis}.
    This quantum process is calculated in light-cone perturbation theory~\cite{Nikolaev:1990ja,Dosch:1996ss}, and is quantified by the light-cone wavefunctions $\psi_{T,L}^{\gamma^* \to q_f \bar q_f}$ of the virtual photon.
    The squared moduli of these virtual photon splitting wavefunctions are:
        \begin{align}
        \abs{\psi^{\gamma^* \to q_f \bar q_f}_T(\rt, Q^2, z)}^2 = & \frac{2 \nc \alpha_{\mathrm{em}} e_f^2}{\pi}
        \Big( m_f^2 \besk_0^2(\varepsilon_f r) 
        \notag\\ \label{eq:phot-split-wavef1}
        & \!\! + \big[z^2 + (1- z)^2\big] \varepsilon_f^2 \besk_1^2(\varepsilon_f r) \Big), 
        \\
        \abs{\psi^{\gamma^* \to q_f \bar q_f}_L(\rt, Q^2, z)}^2 
        = &
        \frac{8 \nc \alpha_{\mathrm{em}} e_f^2}{\pi}
        Q^2 z^2 (1-z)^2 \mathrm{K}_0^2(\varepsilon_f r), \label{eq:phot-split-wavef2}
    \end{align}
    where $\nc$ is the number of color charges in QCD, $e_f$ is the fractional charge of quark flavor $f$, and $m_f$ its mass. The shorthand notation used is $\varepsilon^2_f = z(1-z)Q^2 + m_f^2$, and $\mathrm{K}_i$ are Bessel functions of the second kind.

    The second piece of the equation~\eqref{eq:dis-cross-section} we need to elaborate on is the dipole--target scattering cross section $\sigma_{q \bar q}$ which quantifies the probability of the quark-antiquark dipole state interacting with the target.
    In general this dipole cross section depends on where the quark and antiquark hit the target in the transverse plane. However, typically simplifying assumptions are made, not least because the measured inclusive scattering cross section averages over all target configurations, which does not allow for resolution into the transverse structure of the target. In the simplest case, the target is assumed to be a rotation symmetric density, and the dipole scattering cross section only depends on the size of the dipole-state $r$, and the effective transverse area of the target $\frac{\sigma_0}{2}$. With these assumptions the dipole scattering cross section becomes:
    \begin{equation} \label{eq:dipole}
        \sigma_{q \bar q}(r,x) =
        2 \int_{\mathbb{R}^2} N(\rt,\bt,x) \ud^2 \bt
        \approx
        \sigma_0 N(r,x),
    \end{equation}
    where the impact parameter $\bt$ is the transverse separation between the probe and target, and $N$ is the dipole amplitude~\cite{Gelis:2010nm,Albacete:2014fwa}, which now only depends on the relative distance between the quark and antiquark $r$ and the Bjorken-$x$.
    The dipole amplitude $N(r,x)$ is the unknown function describing the scattering which we want to reconstruct from the data\footnote{Specifically, we approximate that the impact parameter $\bt$ dependence of the dipole amplitude factorizes, separating the integral over the impact parameter so that it can be calculated independently, if some average shape profile for the target is assumed. With a gaussian profile the integral yields the average area $\frac{\sigma_0}{2}$~\cite{Marquet:2007nf}.}.
    
    Equations~\eqref{eq:dis-cross-section}, \eqref{eq:phot-split-wavef1}, \eqref{eq:phot-split-wavef2}, and \eqref{eq:dipole} form the forward problem taken as the starting point in this work. The forward problem terminology is used in the sense that if the dipole amplitude $N(r,x)$ is known, it is possible to calculate the cross sections $\sigma_{L,T}$ in Eq.~\eqref{eq:dis-cross-section} and therefore make predictions about experimentally measured quantities such as the reduced cross section in Eq.~\eqref{eq:sigmared}.
    However, the color glass condensate effective field theory has been used to push the predictive power of this dipole picture further by proving a theory description for the $x$-dependence of the dipole amplitude $N(r,x)$ in the form of an ODE known as the Balitsky--Kovchegov (BK) equation~\cite{Balitsky:1995ub,Kovchegov:1999ua,Kovchegov:1999yj}.
    With the BK equation, it is sufficient to know the initial dipole amplitude $N(r,x_0)$ at a scale $x=x_0$ to be able to calculate predictions at any smaller scale $x<x_0$. 
    The inference of this initial shape is the \emph{implicit inverse problem}, which has conventionally been solved by fitting a parametrized ansatz to data, as discussed in the introduction.
    Recent theory development is working towards enabling a more general description of the initial scale dipole amplitude by relaxing the assumption of Gaussian distribution of color charges in the target~\cite{Penttala:2025tmp}.

    In the next section we apply the inverse problems framework and mathematical observations to rewrite this implicit inference problem into an explicit problem, which can be solved for the unknown quantity without a fit parametrization.
    To enable this, we rewrite the reduced cross section Eq.~\eqref{eq:sigmared} into an integral transform of the dipole amplitude.

\section{Formulation of the dipole amplitude inverse problem} \label{sec:inverse}
    As a starting point for the formulation of the explicit inverse problem, we take the standard result for the leading order accuracy DIS cross section \eqref{eq:dis-cross-section}, and define a new notation for the $z$-integrated wavefunction kernel:
    \begin{equation} \label{eq:z-int-help}
        \mathcal{Z}_{T,L}(r,Q^2) = \frac{Q^2}{4 \pi \aem} \sum_f \int_0^1
            \abs{\psi^{\gamma^* \to q_f \bar q_f}_{T,L}(r, Q^2, z, f)}^2 \ud z.
    \end{equation}
    With this, and the transition to polar coordinates, we have for the structure functions in Eq.~\eqref{eq:structure-functions}
    \begin{equation}
        F_{T,L} \!(x, Q) =
            \int_{0}^{\infty} \!\!
            \mathcal{Z}_{T,L}(r,Q^2) \,
            \frac{\sigma_0}{2} N(r,x)
            r \ud r,
    \end{equation}
    which is essentially an integral transform of the dipole amplitude $N(r,x)$ against the kernel $r \mathcal{Z}(r,Q^2)$.
    Next we rewrite the formula for the reduced cross section:
    \begin{align}
        \sigma_r(x,Q^2) & = 
        \frac{Q^2}{4 \pi \aem}
            \int_{0}^{\infty} \!\!
            \Big[
                \mathcal{Z}_{T}(r,Q^2)
                \notag \\
                &
                +
                \frac{2(1-y)}{1+(1-y)^2} \mathcal{Z}_{L}(r,Q^2)
            \Big]
            \frac{\sigma_0}{2} N(r,x)
            r \ud r
            \notag \\
            & = 
            \int_{0}^{\infty} \!\!
            \mathcal{Z}(r,Q^2,y)
            \frac{\sigma_0}{2} N(r,x)
            \ud r, \label{eq:sigmared-integral-operator}
    \end{align}
    where we changed the order of integration over $r$ and summation in Eq. \eqref{eq:sigmared}, and defined
    \begin{equation}
        \mathcal{Z}(r,Q^2,y) \coloneqq r \Big[
                \mathcal{Z}_{T}(r,Q^2)
                +
                \frac{2(1-y)}{1+(1-y)^2} \mathcal{Z}_{L}(r,Q^2)
            \Big].
    \end{equation}
    At a fixed Bjorken-$x$, the equation~\eqref{eq:sigmared-integral-operator} is now written into a linear integral transformation of the dipole amplitude $N(r)$, which yields the measured quantity $\sigma_r$.
    Various inverse problems are of this form, such as computed tomography, and are computationally solved for the unknown quantity without the necessity to have a functional parametrization to fit to the data.
    \edit{Since we do this computation at a fixed $x$, we are not using any \emph{a priori} information about the $x$-dependence of the solved dipole amplitude. On one hand using an evolution equation like the BK ODE would enable enforcement of smoothness of the reconstructed solutions in the $x$-direction, but on the other hand it would introduce more modeling uncertainty from the accuracy of the applied BK equation, such as modeling of the running coupling in the implementation of the leading order BK equation.
    The freedom from an evolution equation in the present approach can enable the comparison of the reconstructed dipole amplitude with various evolution equations to study in-detail the differences between different phenomenological models of the initial condition, running coupling, and the evolution equation resummation schemes in contrast with the dipole amplitude reconstructed without having to make any of these assumptions in the process. In our view, this novel approach provides an excellent opportunity to study the BK evolution equation and other theory or modeling assumptions in a novel complementary frame, built only on the perturbative dipole picture of DIS and mathematical insight of the inference problem.}

    In contrast with the literature, we associate the dipole amplitude with the overall normalization factor $\frac{\sigma_0}{2}$ of the reduced cross section, which is by definition the average transverse area of the target proton. This means that all the unknown non-perturbative quantities are defined within the functional quantity to be reconstructed from the data, and the forward integral operator is purely defined within light-cone perturbation theory.
    In practice we define this with:
    \begin{equation}
        \frac{\sigma_0}{2} \coloneqq \max_{r >0} \big(N_{\mathrm{rec.}}(r, \xbj) \big),
    \end{equation}
    which means that the reconstructed dipole amplitude does not have to be monotonically increasing at large $r$. This also implies that the transverse area of the target can be reconstructed from the data independently at each Bjorken-$x$. Our aim is to construct the reconstruction process to be general enough to be able to capture that kinds of effects, while keeping the overall normalization incorporated into the definition of the dipole amplitude.
    It is feasible that the dipole amplitude would tail off at large $r$, which is discussed in Ref.~\cite{Mantysaari:2018zdd}, and the goal is that the reconstruction is general enough to enable the quantification of this from real data.
    On the other hand, it is known that the diameter of the proton slowly grows as Bjorken-$x$ decreases~\cite{Caldwell:2009ke}, which we hope to see some evidence of when this reconstruction approach is applied to HERA data.
    
    To enable straightforward numerical implementation, we can discretize the integral transformation in Eq. \eqref{eq:sigmared-integral-operator} to arrive at
    \begin{equation} \label{eq:discrete-sigmar-problem}
        \sigma_{r}^{\mathrm{d}}(x,Q^2) = \sum_{i=0}^{M-1} (r_{i+1}-r_i) \mathcal{Z}(r_i, Q^2, y) \frac{\sigma_0}{2} N(r_i,x),
    \end{equation}
    where $r_i$ are the grid points for $r$. The grid size $M=256$ used in this work was chosen so that the numerical implementation reaches at least $0.1 \%$ relative precision in comparison to the benchmark implementation~\cite{Beuf:2020dxl,Hanninen:2021byo,Hanninen:2022gje}.
    Larger grid size would of course be more precise, but would cause a tradeoff in the stability of the reconstruction, which is more well behaved, when the grid sizes for the data and the reconstruction are at least somewhat proportional.
    Let us then also discretize in $Q$ with the grid $(Q_j)_{j=0}^N$, taking $y_j = y(Q_j^2,x)$ at fixed $x$ and $s$, we can then write the problem as a proper linear equation:
    \begin{equation} \label{eq:linear-dipole-ip}
        \sigma^{\mathrm{d}}_{r,j} \coloneqq \sigma_{r}^{\mathrm{d}}(Q^2_j, x) = \varsigma_{ji} n^i,
    \end{equation}
    where we absorbed $\frac{\sigma_0}{2}$ into $n^i$, and the interval length $(r_{i+1}-r_i)$ into the definition of the discrete reduced cross section forward operator $\varsigma_{ij}$:
    \begin{equation}
        \varsigma_{ij} \coloneqq
        (r_{i+1}-r_i) \mathcal{Z}\big(r_i, Q_j^2, y=y(x,Q_j)\big).
    \end{equation}
    This gives us a tractable discrete linear algebra formulation of the inverse problem, where we can reconstruct the discrete dipole amplitude $n^i$ from a dataset for the discrete reduced cross section $ \sigma^{\mathrm{d}}_{r,j}$ using standard methods for discrete inverse problems~\cite{Hansen:2017airtools}, analogously to applications in image reconstruction and computed tomography~\cite{Hansen:2021ct}.

\section{Numerical reconstruction and results} \label{sec:results}

    With the dipole inverse problem written into the discrete linear form in Eq.~\eqref{eq:linear-dipole-ip}, we can numerically solve for the dipole amplitude from cross section data in a standard software package used for reconstruction problems of this type. 
    We use the Regtools~\cite{Hansen:1994:regtools,Hansen:2007:regtools} and AIR Tools II toolboxes~\cite{Hansen:2017airtools} for their implementations of algebraic iterative reconstruction methods that are used in finding regularized solutions to discrete inverse problems, such as Eq.~\eqref{eq:linear-dipole-ip}.
    In this article we demonstrate a closure test of this reconstruction process, i.e. that if the dipole amplitude is known a priori, we are able to reconstruct it from reduced cross section data computed from that dipole amplitude.
    For this test, we chose the leading order Bayesian inference fits of the dipole amplitude from Ref.~\cite{Casuga:2023dcf} as the so-called \emph{ground-truth}, which is the notion of the precisely known a priori data of the quantity being indirectly measured, and the task is to implement the reconstruction to recover this ground-truth as closely as possible. This development of the reconstruction process is done with the fit parametrization dipole amplitudes, since the true natural ground-truth has never been measured.

    \begin{figure*}[tb]
        \centering
        \includegraphics[width=0.95\textwidth]{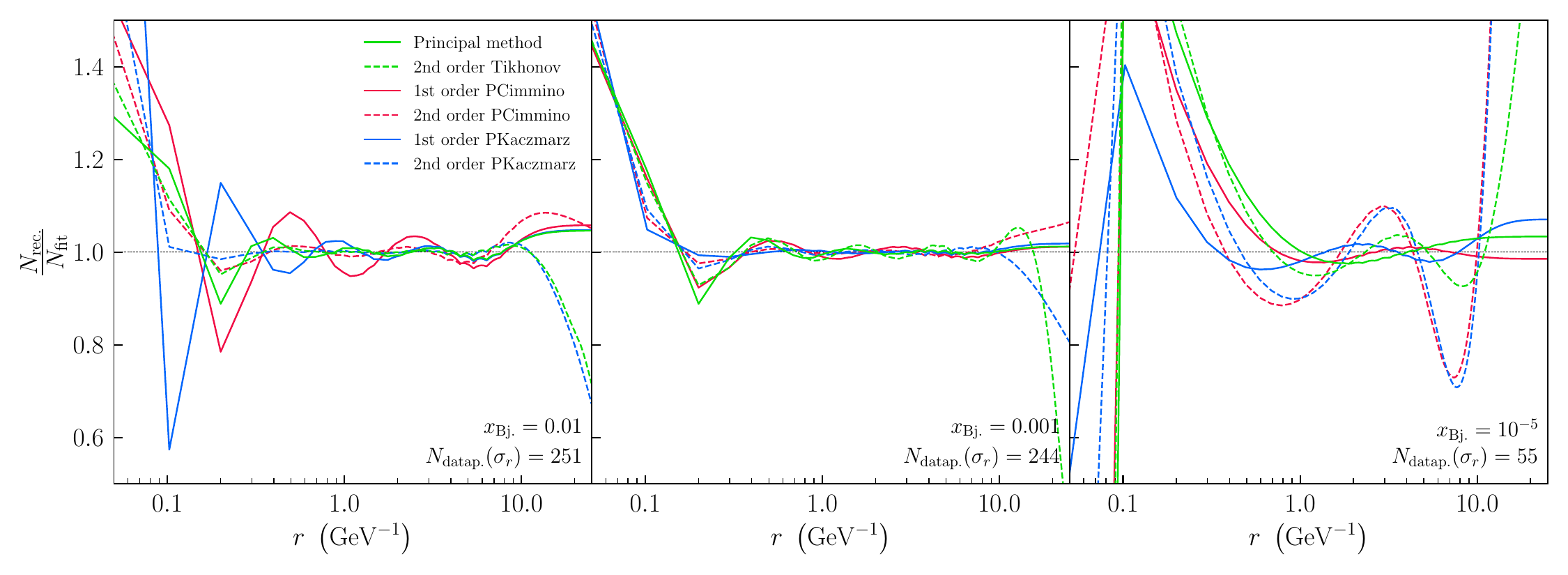}
        \caption{Comparison of considered reconstruction methods.
            Altogether nine algorithms were compared, and the best performing six are shown here.
            The \ordinalnum{1} order priorconditioned Tikhonov--Phillips regularization algorithm was chosen as the main method to be used for the rest of the paper for its consistent good performance, and less pronounced bad behavior in challenging regimes of the reconstruction.
            }
        \label{fig:rec-methods}
    \end{figure*}

    The first step of the implementation is the selection of the specific algorithm used for the regularized reconstruction. Regularization is essentially the mathematical method used in the elimination of over-fitting the data while finding a solution to the inverse problem~\cite{Mueller-Siltanen:2012:ip-book}.
    Different algorithms can be useful for different inverse problems, where the mathematical nature of the inference problem and the quality of the data can have an impact on which is the optimal algorithm to use.
    In this work, all the algorithms that we consider were constructed to efficiently find---at least approximate---solutions to the linear equation
    \begin{equation} \label{eq:linear}
        \mathbf{y} = \mathbf{A} \mathbf{x},
    \end{equation}
    which is the form of the explicit inverse problem we are solving in Eq.~\eqref{eq:linear-dipole-ip}.
    In the case that the matrix $A$ is not invertible---such as when it is not square---approximations such as the Moore--Penrose pseudoinverse~\cite{Hansen:2021ct} $A^\dagger$ of $A$ can be leveraged to iteratively find an approximate solution to the linear system of equations \eqref{eq:linear}.
    A more in-depth and pedagogical exposition of these algorithms is given in Ref.~\cite{Hansen:2021ct}.
    
    We compared altogether 9 algorithms to solve the linear system of equations~\eqref{eq:linear} to determine which performs best for our problem. The unconditioned algorithms we tested are the Tikhonov--Phillips regularization~\cite{Tikhonov:1943,Tikhonov:1963,Tikhonov:1977,Phillips:1962,Hoerl:1962},
    and Cimmino's~\cite{Cimmino:1938} and Kaczmarz's~\cite{Kaczmarz:1937, Gordon:1970:ART} methods, which are \ordinalnum{0} order methods and implemented in Ref.~\cite{Hansen:2017airtools}. We compared these methods with the corresponding \ordinalnum{1} and \ordinalnum{2} order priorconditioned methods, which essentially regularize the solution for its $N^\mathrm{th}$ order derivative instead of its $2-$norm, and were implemented in Ref.~\cite{calvetti2007a,hansen2010a,Schluter:2017iterative}.
    Qualitatively, the \ordinalnum{0} order algorithms seek to minimize the $2-$norm of the reconstructed dipole amplitude at any point while describing the data as well as possible, whereas the higher order priorconditioned methods minimize the derivatives of the respective order of the reconstruction. The rough intuition here might be that if two candidate solutions describe the data nearly identically, but one has larger maxima---or those of its derivatives---than the other, the regularization algorithm prefers the less extreme solution, since the measurement is not sensitive to the large fluctuations.
    
    We found that the standard \ordinalnum{0} order methods performed badly or completely failed to reconstruct the dipole amplitude---for example by finding oscillating solutions or solutions that had negative minima---and to describe the simulated data, so they were eliminated from the comparison.
    The accuracy of the methods that managed to reconstruct the fit dipole amplitudes at least to some degree are compared in Fig.~\ref{fig:rec-methods}.
    The method denoted in Fig.~\ref{fig:rec-methods} as the ''principal method`` was selected as the main method to go forward with the closure test implementation, and is a slight variation of the \ordinalnum{1} order Tikhonov--Phillips method. It was selected for recovering the ground-truth fit dipole most accurately on average, although in some cases other methods could perform as well or slightly better, so the selection was not quite obvious. It was consistently among the best performing methods in the intermediate-$r$ regime, while being among the least badly behaving in the high and low-$r$ regimes. The two other tested \ordinalnum{1} order methods also performed fairly well in some situations, which perhaps suggests that the \ordinalnum{1} order methods are enforcing the right type of smoothness of the solutions.
    
    The 1st order Tikhonov--Phillips regularization chosen as the principal method for the closure test optimizes the following function when computing the reconstructed solution:
    \edit{
    \begin{equation} \label{eq:tikhonov-1st}
        \underset{\mathbf{n}\in \mathbb{R}^M}{\argmin} \, \lbrace || \varsigma \mathbf{n} - \sigma_r ||_2^2 + \lambda || \frac{\ud}{\ud r} \mathbf{n} ||_2^2 \rbrace,
    \end{equation}
    }
    where $\mathbf{n}$ is the reconstructed dipole amplitude, ${\argmin}$ indicates that the expression is minimized for $\mathbf{n} \in \mathbb{R}^M$ where $M$ is the number is discretization points, $\lambda$ is a regularization weight parameter, and the $\ell^2$-norm is defined for a real vector $\mathbf{x}$ as $||\mathbf{x}||_2 \coloneqq \sqrt{\mathbf{x} \cdot \mathbf{x}} \equiv \sqrt{\sum_i^n x_i^2}$. The numerical derivative of the reconstruction is computed using a finite difference approximation of the derivative. The $\lambda$ parameter in the 1st order Tikhonov--Phillips method limits how large the first derivative of the reconstructed dipole amplitude can be, with large $\lambda$ corresponding to a large weight against high values of the derivative, and inversely small $\lambda$ implements only a small weight for the value of the derivative. In practice this means that with large $\lambda$ the reconstructed solutions are ''stiff`` and very smooth, and tend to under-fit the data, and conversely allowing for too small $\lambda$ leads to strongly fluctuating reconstructions that try to over-fit the data. In the case of simulated data with no noise, the over-fitting is less of an issue, and is harder to detect, but when we implement the confidence interval sampling for the reconstructions, it becomes very obvious when the reconstruction tries to over-fit to the noise in the data.
    We implement and test two variations of the regularization method: one in ideal conditions with no statistical noise in the data to reconstruct the ground-truth as precisely as possible, and another motivated by the application to real data with statistical uncertainties.

    An essential part of the regularization and reconstruction process is to quantify the viability and accuracy of each candidate solution in the space of all feasible dipole amplitudes. This is done with an error function, which is then minimized analogously to a standard fitting process, and which can for example give a larger error to a candidate solution that has unphysical properties. For the ''noiseless`` data reconstruction we use the error function defined as:
    \begin{equation} \label{eq:err-noiseless}
        \mathrm{err}_{\mathrm{noiseless}}(\mathbf{n}) \coloneqq \frac{\norm{\sigma_{r} - \varsigma \mathbf{n}}_2}{\norm{\sigma_{r}}_2},
        + \varepsilon_{-} \theta(- \min (\mathbf{n}))
    \end{equation}
    which seeks to minimize the relative error of the reconstruction $\mathbf{n}$ to each data point\edit{, and the second term with $\varepsilon_{-} = 10^{-3}$ penalizes reconstructions with non-positive minima.}
    We treat the regularization parameter $\lambda$ as a nuisance parameter defined on the interval $\lambda \in [ 5 \cdot 10^{-4}, 9.5 \cdot 10^{-2}]$.
    The second variant of the regularization constructed for application to real data uses an error function based on the $\chi^2$ test, which assumes that the measurement has statistical uncertainty:
    \begin{align}
        \mathrm{err}_{\mathrm{\chi^2}}(\mathbf{n})
        \coloneqq& \left| \frac{\chi^2(\mathbf{n})}{N-1} - 1 \right|
        \\
        =& \left| \frac{1}{N-1} \sum_{i=1}^N\frac{(\sigma_{r,i} - (\varsigma \mathbf{n})_i)^2}{\delta_i^2} - 1 \right|,
        \label{eq:err-chisq}
    \end{align}
    where $\delta_i$ is the absolute error of the measurement $\sigma_{r,i}$.
    This error function is used to find the $\lambda$ and the corresponding reconstruction $\mathbf{n}$, whose agreement with the noisy data is closest to $\frac{\chi^2}{N-1}=1$. 
    \edit{In practice this means that to each noisy dataset we minimize the error \eqref{eq:err-chisq} to find an optimal $\lambda$ and the corresponding best candidate for the reconstruction $\mathbf{n}_\lambda$. This can be seen as fitting a free nuisance parameter to the noisy dataset, where any $\lambda$ results in a candidate reconstruction through the regularization algorithm \eqref{eq:tikhonov-1st}, and the agreement of said candidate is quantified by the error \eqref{eq:err-chisq}. Optimal $\lambda$ is then the value whose corresponding reconstruction best agrees with the data with the $\chi^2$-test.
    The noiseless error is used similarly, but with the assumption that the simulated data does not have any noise, and so the algorithm can optimize for a much closer agreement with the noiseless data. The non-positivity penalty term in Eq. \eqref{eq:err-noiseless} acts indirectly to limit overfitting by preventing non-physical solutions that have negative minima that would arise when overfitting noise in the data.
    We use the error functions \eqref{eq:err-noiseless}, \eqref{eq:err-chisq} to implement so-called unsupervised regularization, where the regularization method fixes the parameter $\lambda$ without precision knowledge of the ground-truth. The alternative of supervised regularization would use the ground-truth fit parametrization to tune the regularization algorithm, but we do not wish to train the regularization for a fixed model of the dipole amplitude, such as the MV model, and instead aim for the reconstruction implementation to be general enough to be able to differentiate between different models, and to enable comparison of the real data reconstruction to these pre-existing models. More rigorous methods of unsupervised regularization have been developed such as the discrepancy principle~\cite{Engl:1987:discrepancy-principle}, L-curve~\cite{Hansen:2000:L-curve} and Generalized Cross-Validation~\cite{Hansen:1994:regtools,Golub:1979:generalized-cross-validation-gcv}, and would be desirable to implement for a more robust reconstruction from HERA data.} The behavior of these noisy reconstructions are quantified in the rest of this article by showing a random representative of a reconstruction to a dataset with random noise, and the point-wise mean of all the reconstructions performed to sets of randomly sampled data sets.

    We perform the closure test as follows: first with a chosen fit parametrization of the dipole amplitude, we compute ''simulated`` reduced cross section data with fixed $\sqrt{s} = 318.1 \, \mathrm{GeV}$ at various fixed values of Bjorken-$x$ over a range $0.25 \gev^2 \leq Q^2 \leq 150 \gev^2$. This computation is performed using the implementation used in Refs.~\cite{Ducloue:2017ftk,Hanninen:2017ddy,Beuf:2020dxl,Hanninen:2022gje} to separate the generation of the reduced cross section data from the implementation of the reconstruction~\footnote{Using the same discretized implementation of the inverse problem for both data generation and reconstruction can produce results that are ''too good``, a phenomenon known as ''inverse crime``~\cite{Mueller-Siltanen:2012:ip-book,wirgin2004inversecrime}.}.
    The reconstruction of the dipole amplitude from the "simulated`` reduced cross section data proceeds in two stages: first we compute the discretized forward operator using a uniform grid of $256$ points for $r \in [0.005, 25] \, \mathrm{GeV}^{-1}$ with fixed $\sqrt{s}$, light quark mass $m_{\mathrm{light}} = 0.14 \, \mathrm{GeV}$, and charm quark mass $m_c = 1.35 \, \mathrm{GeV}$. This means that the forward operator is completely static, and does not need to change during the reconstruction process, if we assume that the quark masses are fixed. This significantly speeds up the computation, which as linear algebra operations are very fast.

        \begin{figure*}[thbp]
            \centering
            \includegraphics[width=0.95\textwidth]{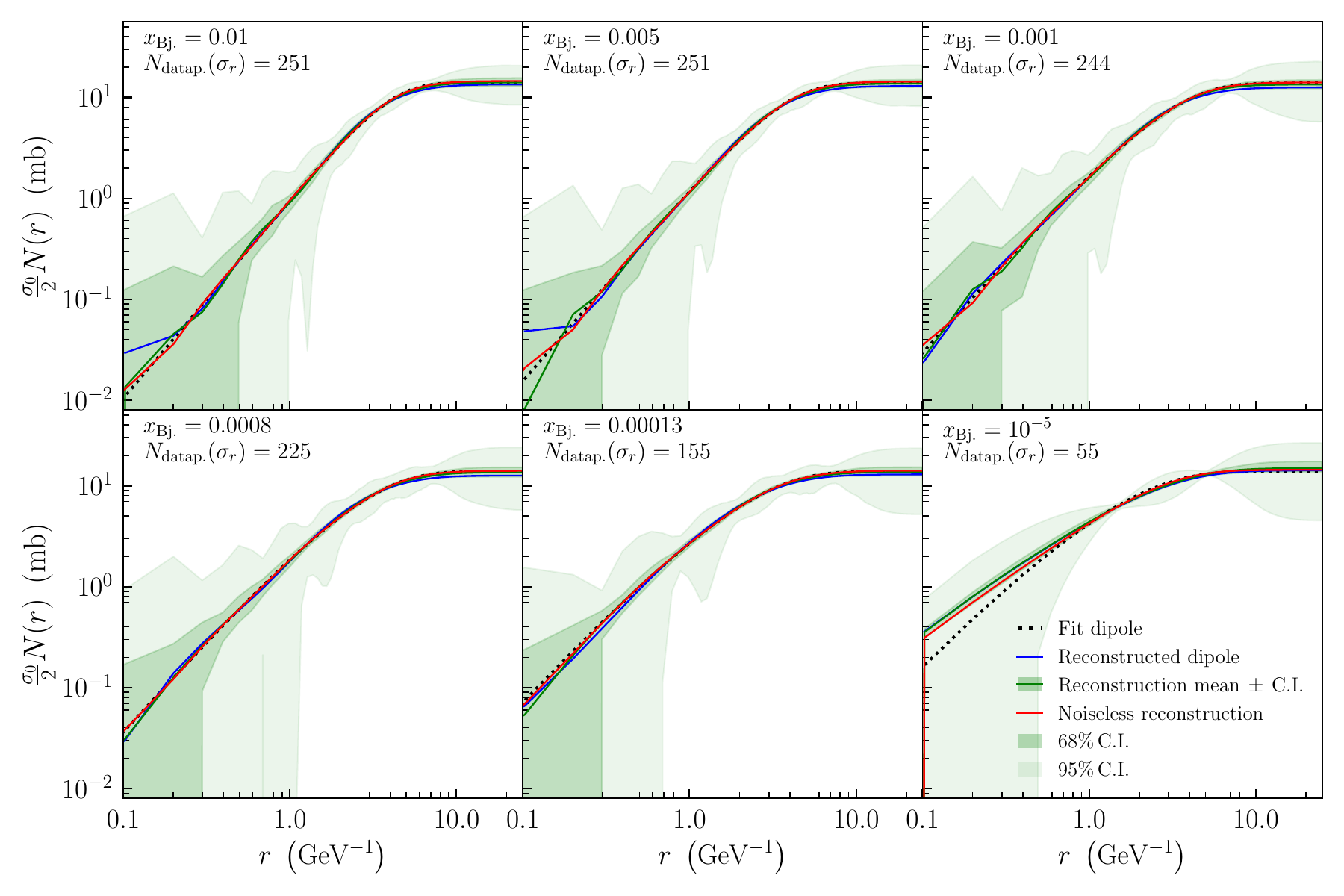}
            \caption{Reconstructions of the dipole amplitude at various fixed Bjorken-$x$ from generated reduced cross section data compared to the ground-truth dipole amplitude. The ground-truth dipole amplitude is the 4-parameter Bayesian fit from Ref.~\cite{Casuga:2023dcf}.}
            \label{fig:plot1-dip-light}
        \end{figure*}
    
    The second stage performs the actual computation to solve the inverse problem for the dipole amplitude, and proceeds in two steps.
    First we reconstruct directly to the simulated reduced cross section data points with no error, which we call the ''noiseless`` reconstruction, and is the best estimate for the dipole amplitude as implied by the data. Ideal here would be to perfectly recover the fit dipole amplitude functional shape.
    However, the real data will have experimental uncertainty that will limit the accuracy and confidence of the reconstruction. To quantify this, we perform a second step of the reconstruction to gauge the uncertainty associated with the reconstructed dipole amplitude if the data has some defined uncertainty. We assume a normally distributed relative uncertainty of $1 \%$ for the generated reduced cross section, which is of the same order of magnitude as the error in the real data~\cite{Abramowicz:2015mha, H1:2018flt}, and sample $N_{\mathrm{sample}}=10^3$ representatives of the noisy dataset from the distributions of each data point. We then run the reconstruction using the error function \eqref{eq:err-chisq} to each of these noisy samples of the $\sigma_r$, recording reach reconstruction $N_{m}^{\mathrm{noisy}}$, and the corresponding reduced cross section from that dipole $\sigma_r^m = \tilde{Z} N_{m}^{\mathrm{noisy}}$. This enables us to define the point-wise distributions of the noisy reconstructions and the reduced cross sections predicted by those dipoles, from which we can calculate the confidence intervals of the reconstructions at each grid point $r_i$, and of the predicted $\sigma_r$ at $Q_i$. The mean and confidence intervals of $68\%$ and $95\%$ are shown wherever applicable, and further discussed below with the result analysis. For normally distributed uncertainties these would roughly correspond to one and two units of standard deviation, but we observed that the point-wise distributions of the noisy reconstructions do not always obey the normal distribution. For example, at small $r$, where the reconstruction is dissuaded to take negative values for the dipole amplitude, the distributions can be asymmetrical or multimodal.

        \begin{figure*}[thbp]
            \centering
            \includegraphics[width=0.95\textwidth]{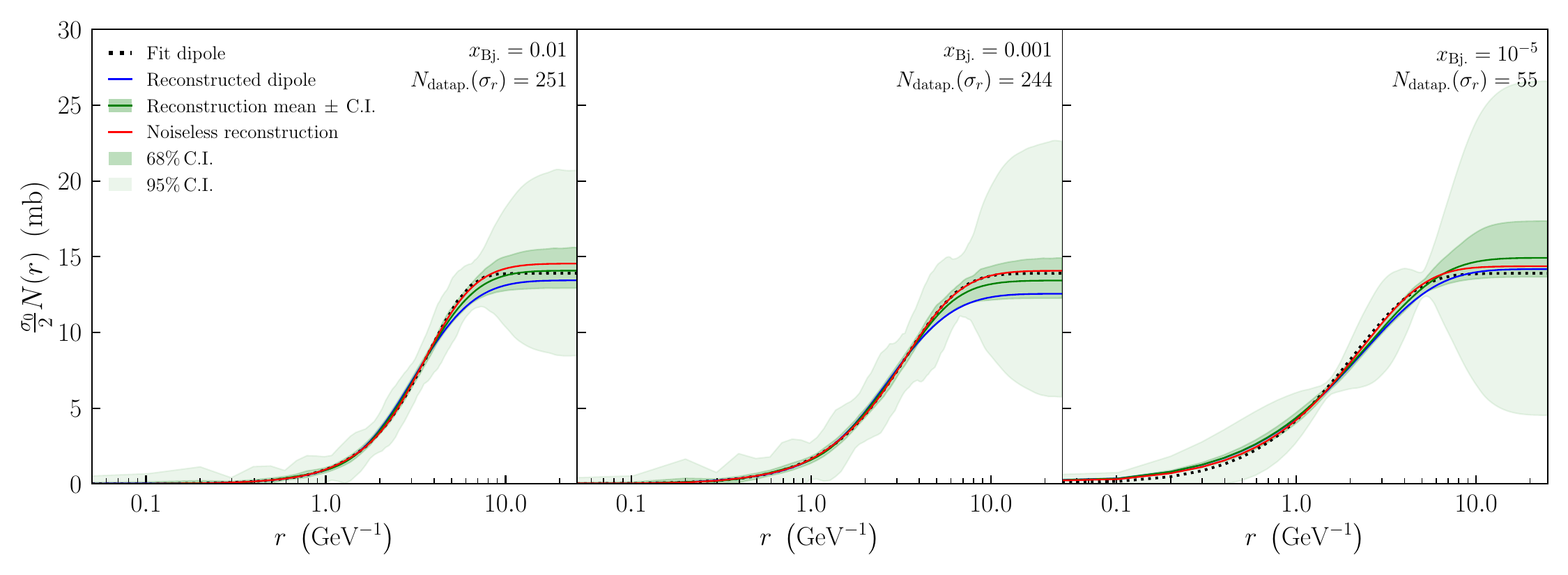}
            \caption{Reconstructions of the dipole amplitude shown with linear vertical axis to show the growth of uncertainty at large dipole sizes, where the reconstruction becomes unsensitive to large changes in the dipole, due to the asymptotic vanishing of the forward operator at large $r$. Nevertheless, the reconstructions manage to recover the ground-truth fairly accurately even at large $r$.}
            \label{fig:plot1b-dip-light-loglinear}
        \end{figure*}
    
        \begin{figure*}[thbp]
            \centering
            \includegraphics[width=0.95\textwidth]{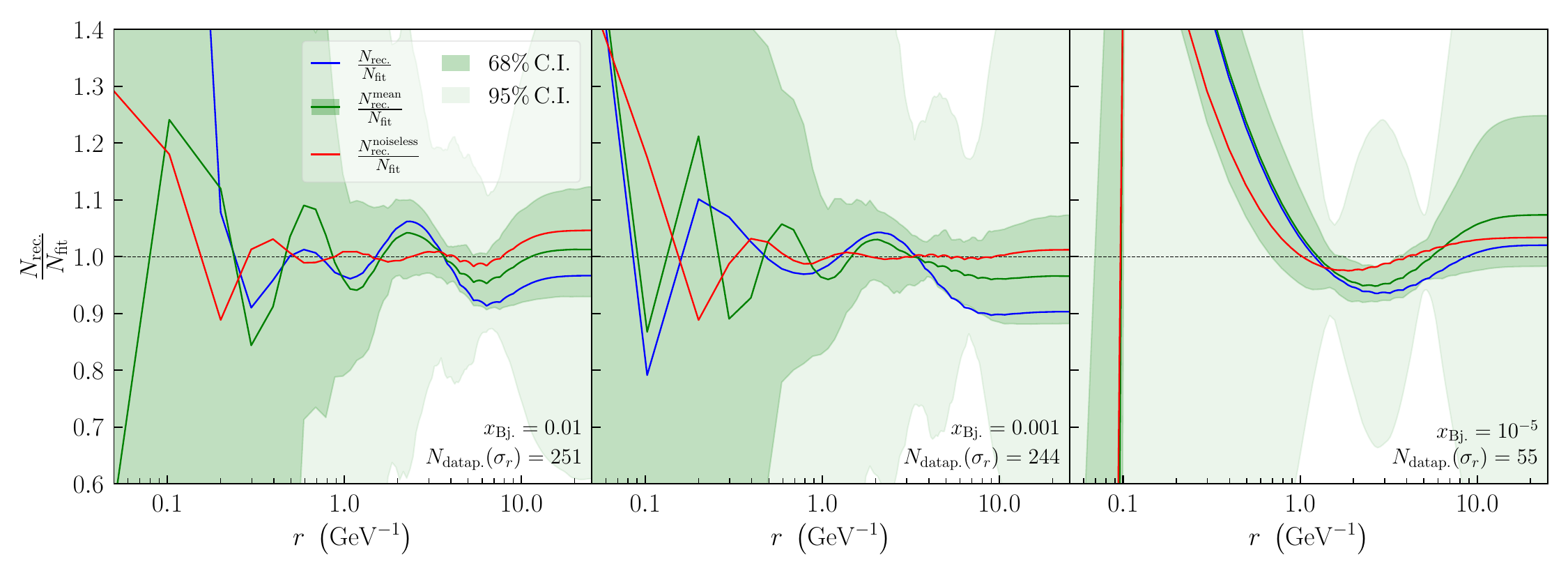}
            \caption{Ratio of the reconstructions with respect to the ground-truth fit dipole amplitude to show the relative precision of the reconstructions. In ideal conditions the ''noiseless`` reconstruction recovers the ground-truth quite accurately, however with the statistical fluctuations introduced to the data, the reconstruction becomes less accurate, and the relative uncertainties grow especially at small and large $r$.}
            \label{fig:plot1b-dip-light-ratio}
        \end{figure*}
    
        \begin{figure}[thb]
            \centering
            \includegraphics[width=0.95\columnwidth]{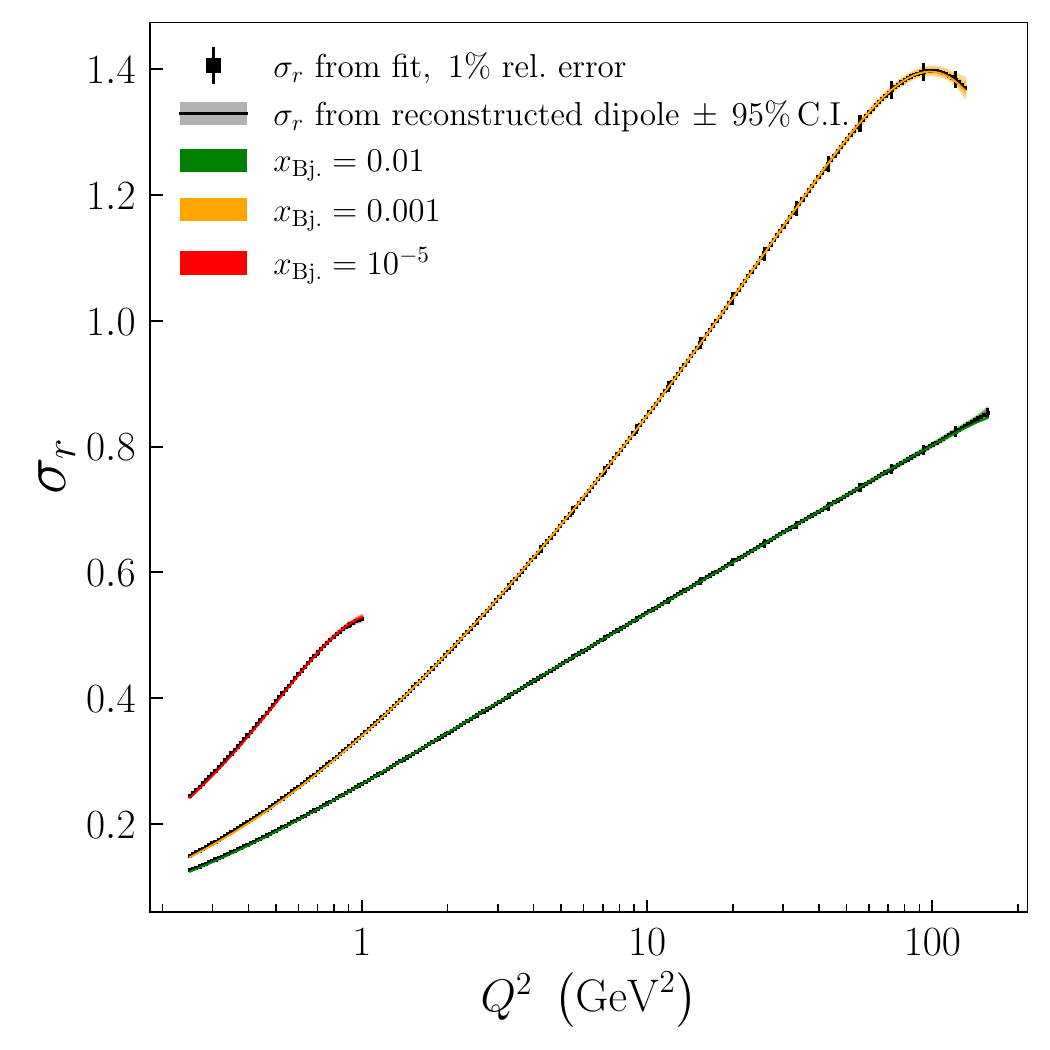}
            \caption{ Comparison of the reduced cross section calculated from the 4-parameter Bayesian fit~\cite{Casuga:2023dcf}, and the calculation from the reconstructed dipoles shown in Fig.~\ref{fig:plot1-dip-light}, including $68\%$ and $95\%$ confidence intervals. Sampling of the reconstruction confidence intervals assumes $1\%$ relative error for the reduced cross section data---which is on the scale of precision of the real data~\cite{Abramowicz:2015mha,H1:2018flt}---with every tenth error bar shown for clarity.}
            \label{fig:plot2-sigma-light}
        \end{figure}

We first show the reconstruction closure test for the 4-parameter Bayesian inference dipole amplitude~\cite{Casuga:2023dcf} in Figure~\ref{fig:plot1-dip-light}. The figure shows the reconstruction performed at various fixed Bjorken-$x$, ranging from $10^{-5}$ to $10^{-2}$, and overall the accuracy of the reconstruction is fairly good. We see that, as expected, at small $r$, where the absolute value of the dipole amplitude is asymptotically vanishing, the uncertainty of the reconstruction grows significantly, while the noiseless reconstruction and the mean of noisy reconstructions still manage to mostly match the ground-truth fit dipole down to $r=0.1 \gev^{-1}$. On the other hand, at large $r \gtrsim 10 \gev^{-1}$, the uncertainty grows as well, shown also in the Fig.~\ref{fig:plot1b-dip-light-loglinear} with linear vertical axis. This increase of the uncertainty is understood by the asymptotic vanishing of the forward operator at large $r$, which stems from the exponentially vanishing Bessel $\besk_{0,1}$ functions. The reconstruction performs best in the intermediate regime, where neither of these effects are hindering the reconstruction process. In Fig.~\ref{fig:plot1b-dip-light-ratio} we show the relative magnitude of the uncertainties as a ratio of the reconstruction and its uncertainties with respect to the ground-truth fit. In this plot we see more clearly the effect of the number of $\sigma_r$ datapoints that are available for the reconstruction: at $\xbj=10^{-5}$, where there are notably fewer datapoints, the reconstruction is less accurate, and its uncertainties grow substantially faster compared to the other cases. The number of datapoints varies, because the reconstruction is done at fixed $\sqrt{s}$ and $\xbj$, with $y<1$, which limits the available range for $Q^2$:
\begin{equation}
    Q^2 = s \xbj y < s \xbj.
\end{equation}
Here we see that as $\xbj$ decreases, so does the upper limit for $Q^2$ for the datapoints $\sigma_r(Q^2,\xbj)$ that are used for the reconstruction. For example, at the smallest $\xbj=10^{-5}$ used in this work this leads to the quite restrictive upper limit $Q^2 \lesssim 1.01 \gev^2$.

Figure~\ref{fig:plot2-sigma-light} shows the reduced cross section $\sigma_r$ data generated from the ground-truth fit dipole amplitude, and for comparison the reduced cross section calculated from the reconstructed dipole amplitude. The $\sigma_r$ with solid colored line is calculated from the noiseless reconstruction, whereas the confidence intervals are calculated from each of the noisy reconstructions done in the process of sampling the reconstruction uncertainties. Each of the noisy reconstructions is used to compute the corresponding reduced cross section, and these are stored. After all the $N_{\mathrm{sample}}$ reconstructions have been run, the point-wise distributions of all the corresponding reduced cross sections are calculated, giving the confidence intervals shown in Fig.~\ref{fig:plot2-sigma-light}. The uncertainties are small, as they should, since each reconstruction is done to a random sample of the dataset within the assumed $1\%$ relative errors, and by construction, the reduced cross section computed from any of the reconstructions has to fall within the relative uncertainty of the generated data. This is in stark contrast with the substantial uncertainties for the dipole amplitudes shown in Figs.~\ref{fig:plot1-dip-light}, \ref{fig:plot1b-dip-light-loglinear}, and \ref{fig:plot1b-dip-light-ratio}, showing the uncertainty in determining the dipole amplitude from inclusive DIS data, especially at small and large dipole diameter~$r$.

From Figs.~\ref{fig:plot1b-dip-light-loglinear} and \ref{fig:plot1b-dip-light-ratio} we can see that the reconstruction of $\frac{\sigma_0}{2}$ works fairly well, with the leading edge of the peak of the dipole amplitude being reconstructed quite accurately, especially in the ideal noise-free conditions, and the relative precision of the reconstruction of the maximum with respect to the ground-truth fit being in the $1\% - 5 \%$ range.
While the variance of the maxima of the reconstructions from random fluctuations in the data is fairly large, the mean of the reconstructions recovers the maximum quite accurately even at large $r$---considering the $1\%$ relative error---since some of the fluctuations are canceled out in the mean.
This growth of the variance at large $r$ stems from the exponentially vanishing forward operator.
Ideally we would like to reconstruct the maximum $\frac{\sigma_0}{2}$ from real reduced cross section data measured at HERA~\cite{Aaron:2009aa,Abramowicz:2015mha,H1:2018flt}, but it seems likely that experimental uncertainties will make that challenging with only inclusive DIS data.
This goal is perhaps more realistically approached as a multi-modal inverse problem, where inclusive diffractive DIS~\cite{Barone:2002cv} data~\cite{H1:2012xlc:ddis} would be incorporated into the reconstruction procedure simultaneously, while avoiding introducing additional modeling uncertainty by including vector meson production data, for example.
    
        \begin{figure*}[thbp]
            \centering
            \includegraphics[width=0.95\textwidth]{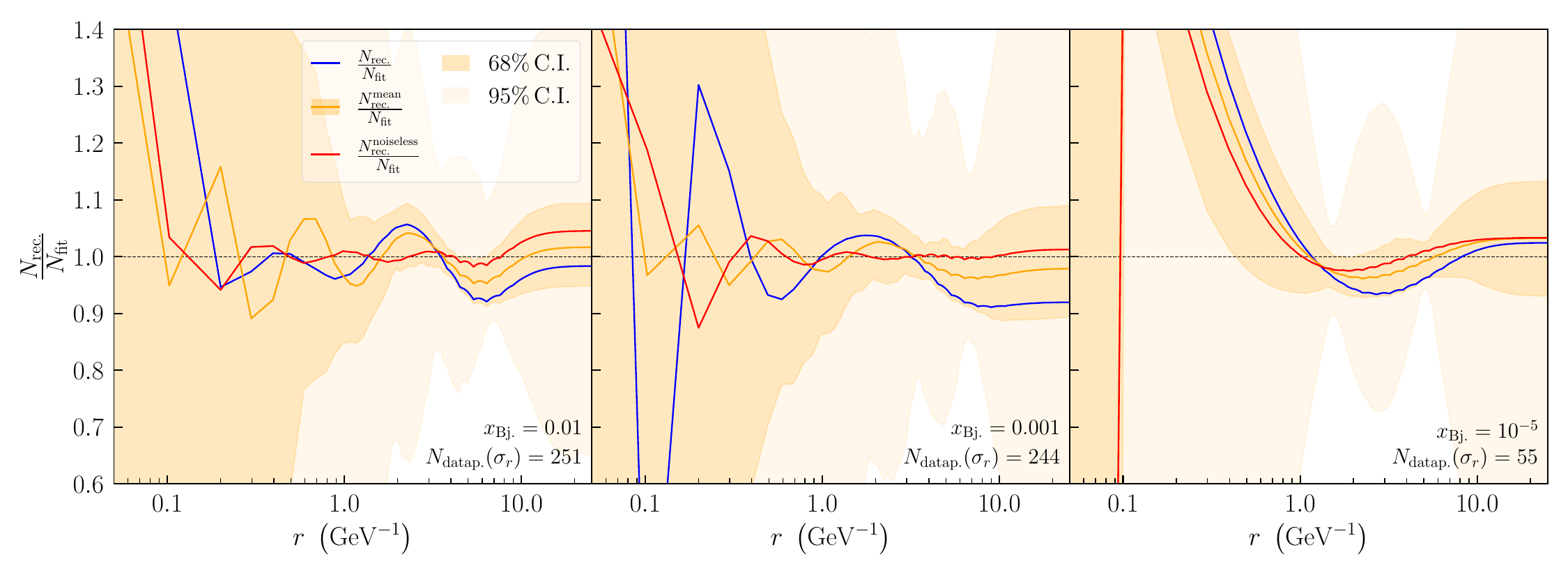}
            \caption{Reconstruction of the dipole amplitude from light-quark-only reduced cross section data calculated from the 5-parameter Bayesian fit from Ref.~\cite{Casuga:2023dcf}.}
            \label{fig:rec-dip-light-5par}
        \end{figure*}
        \begin{figure*}[thbp]
            \centering
              \includegraphics[width=0.95\textwidth]{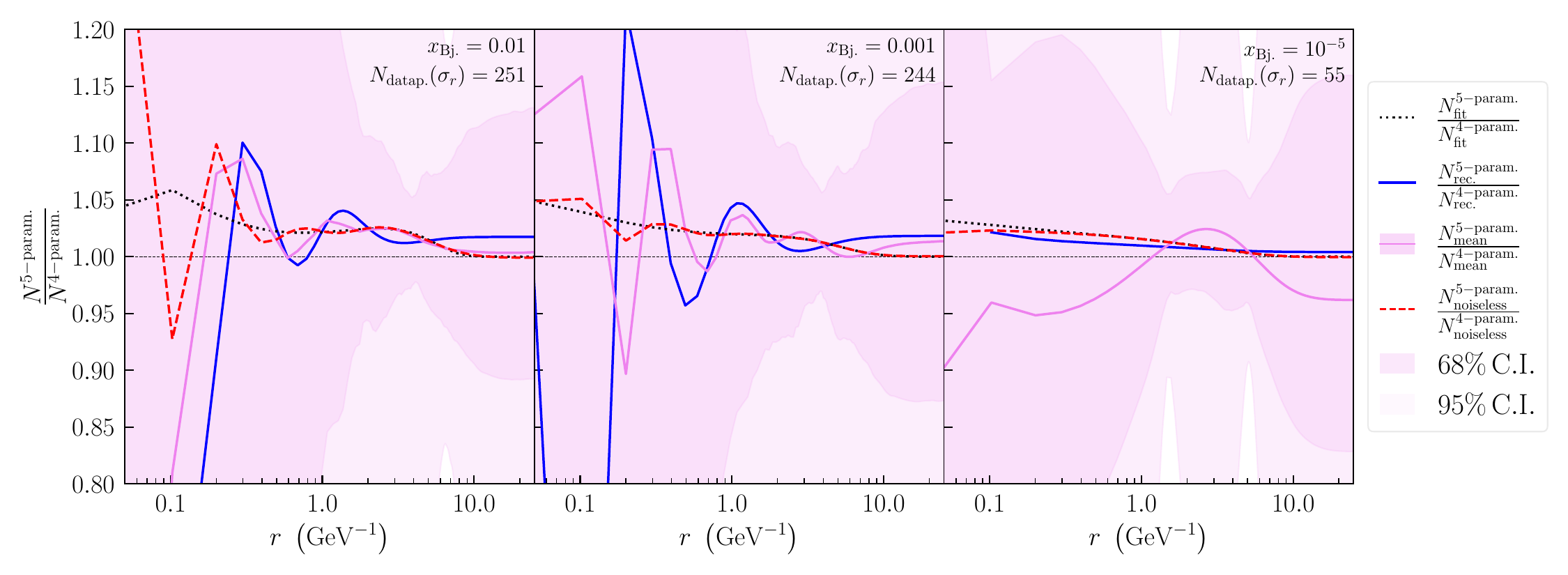}
            \caption{Ratio of the 5-parameter and 4-parameter fit dipole amplitudes compared with the corresponding ratio of their reconstructions. This plot demonstrates that---at least theoretically---the reconstruction process is capable of resolving between the two fit parametrization dipole amplitudes by showing that the reconstructions reproduce the ratio of the parametrizations. 
            The large fluctuations seen at small $r$ are noise in the reconstructions, which is amplified in the ratio of the reconstructions, and a sign that the reconstruction algorithm is diminishing in accuracy in that regime.}
            \label{fig:fit-ratio-vs-rec}
        \end{figure*}
        \begin{figure*}[thbp]
            \centering
            \includegraphics[width=0.95\textwidth]{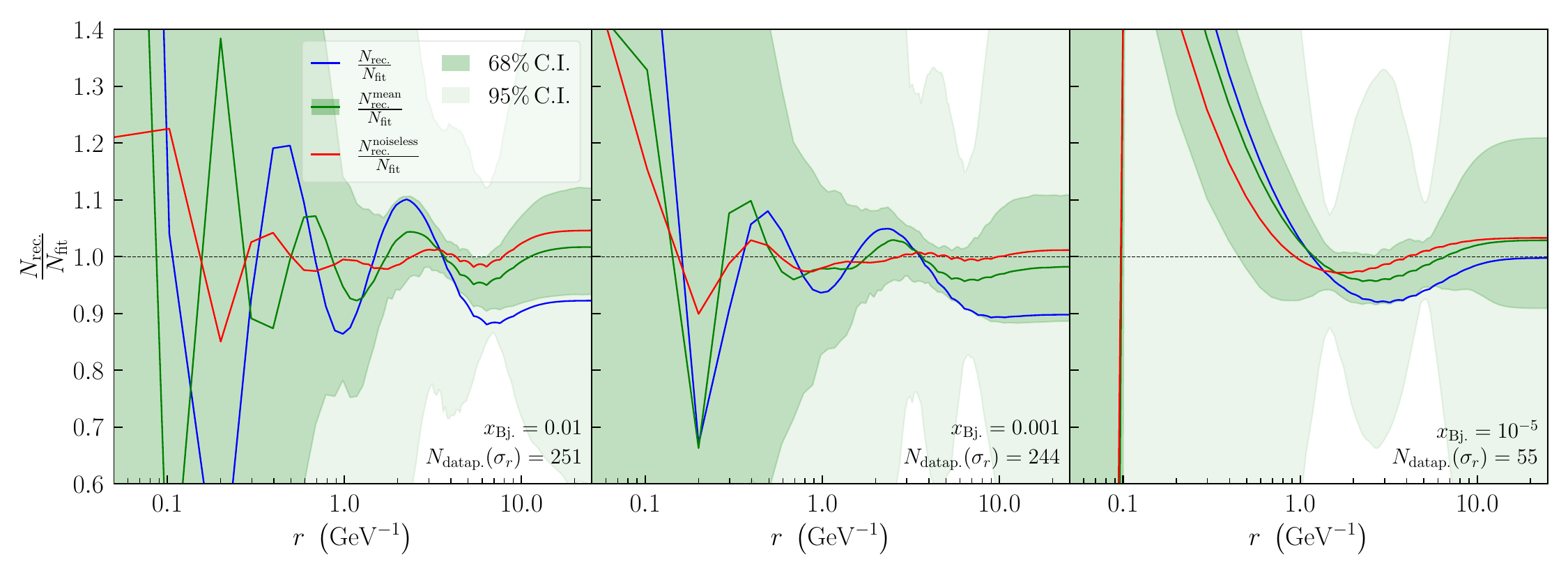}
            \caption{Reconstruction of the dipole amplitude from reduced cross section data using the forward operator including the charm quark contribution. The reference fit dipole is the 4-parameter fit from Ref.~\cite{Casuga:2023dcf}, where charm production is not included. The point of this reconstruction is to show that the inclusion of heavy quarks such as charm in the forward operator does not qualitatively change the inverse problem and that the same methodology applies.}
            \label{fig:rec-dip-charm}
        \end{figure*}

In Fig.~\ref{fig:rec-dip-light-5par} we show the reconstruction of the 5-parameter fit dipole amplitude---which introduces an additional free parameter in the fit~\cite{Casuga:2023dcf}---from the corresponding reduced cross section data. Overall the performance of the reconstruction is very similar as with the 4-parameter fit discussed above, which is expected since the reconstruction process is completely unaware of the fit parametrization used in the computation of the reduced cross section data.
To verify that the reconstruction process is capable of distinguishing between the two slight variations of the dipole amplitude parametrization used in the Bayesian inference, we show in Fig.~\ref{fig:fit-ratio-vs-rec} the ratio of the two dipole amplitude parametrizations.
Even with the discrepancy between the dipole amplitudes being in the single percentage points, in the ideal noiseless conditions the ratio of the two reconstructions is able to very accurately match the ratio of the fit parametrizations, aside of the small-$r$ regime, where the reconstruction becomes inaccurate and unstable.
However, the mean of the reconstructions has more trouble recovering the ratio accurately, especially with the lower number of reduced cross section data points at small Bjorken-$x$, with the best and quite accurate mean reconstruction being in the largest bin with $\xbj=0.01$. 
The noisy reconstruction---shown in blue in Fig.~\ref{fig:fit-ratio-vs-rec}---fails to achieve meaningful accuracy of the ratio, and some systematic effect in the smallest $\xbj=10^{-5}$ bin prevents successful reconstruction of the ratio by the mean.
This capability to quantitatively resolve the fit parametrizations in the meaningful dipole size regime will be crucial for the application of this inverse problems framework to real data, and in theoretical conditions it can be achieved, but experimental uncertainties seem to introduce further challenges.
        
        \begin{figure}[ht]
            \centering
            \includegraphics[width=\columnwidth]{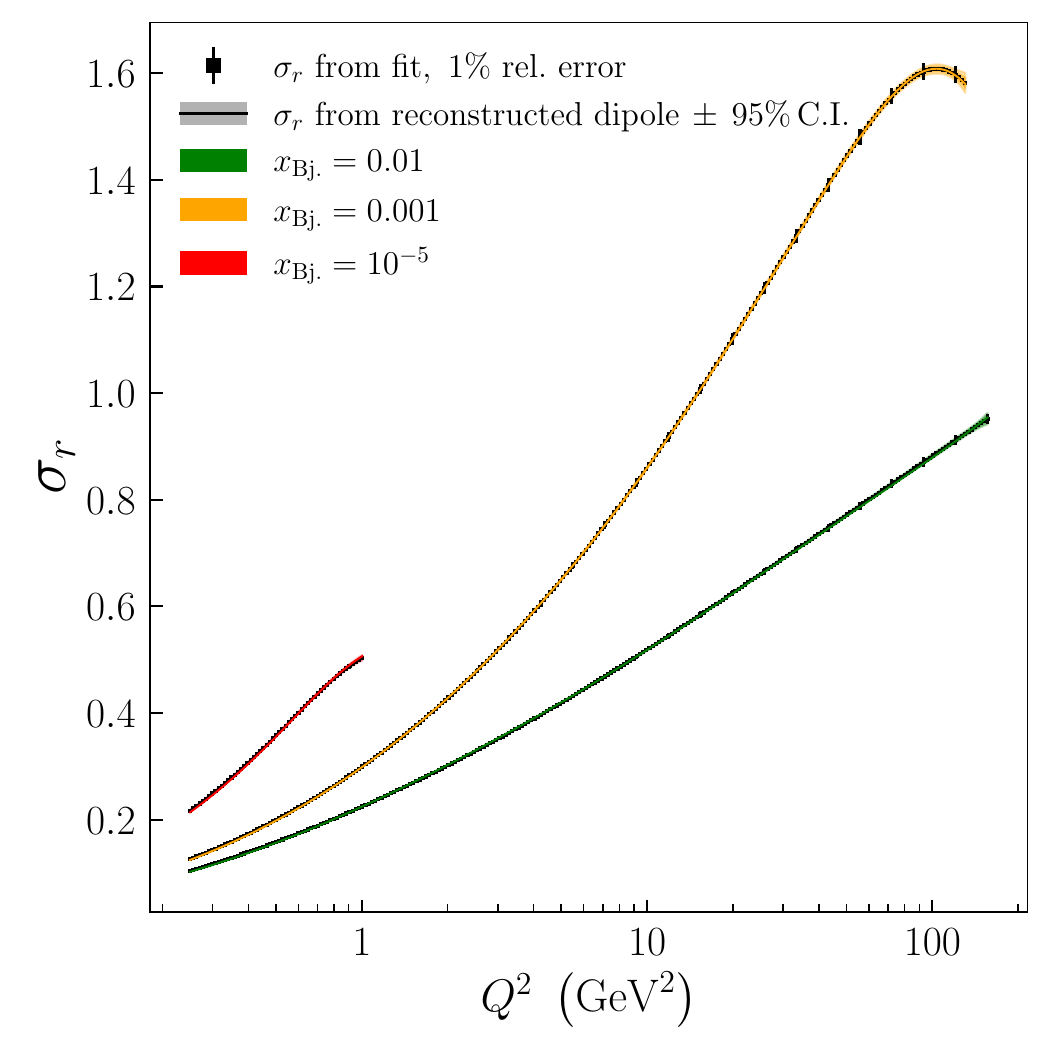}
            \caption{Reduced cross section with charm quark included calculated from the reconstructed dipole amplitudes compared to the calculation from the 4-parameter fit. Every tenth error bar is shown.}
            \label{fig:rec-sigma-charm}
        \end{figure}

Figure \ref{fig:rec-dip-charm} shows that, as expected, the reconstruction works equally well when the contribution of the charm quark is included in the forward operator in Eqs.~\eqref{eq:z-int-help} and \eqref{eq:discrete-sigmar-problem}. Inclusion of the charm contribution is a minor change to the forward operator, which does not affect the reconstruction process, which is seen as a very similar performance of the reconstruction in Fig.~\ref{fig:rec-dip-charm} as was discussed with the light-quark-only reconstruction above.
Fig. \ref{fig:rec-sigma-charm} shows the reduced cross sections computed from the reconstructed dipole amplitudes compared to the predicted cross sections from the fit parametrization dipoles, and as with the light-only case, by definition the variance of the reconstructions must fall within the assumed $1\%$ relative errors of the generated reduced cross section data used in the sampling of the noisy reconstructions.

In addition to the effects contributing to the uncertainty of the reconstruction that have been discussed above we have identified a few other sources: 
perturbation theory precision of the forward operator,
and details of the numerical implementation used such as the selection of $\lambda$ and its viable regime,
and the choice of the 1st order Tikhonov--Phillips regularization method.
This formulation of the inverse dipole problem uses only the leading order light-cone perturbation theory description of the scattering process, which yields inferior theory precision compared to state-of-the-art next-to-leading order results. On the other hand, all of the theoretical contributions are incorporated in the reconstructed dipole, to all orders of the perturbation theory, since the reconstruction yields the dipole amplitude from the data precisely by inverting specifically only the leading order contribution. This implies that once the reconstruction is done from real data, quantified comparisons of the reconstruction can be done against the leading order fit dipole amplitudes to quantify beyond leading order effects.
Furthermore, implementing the forward operator at NLO accuracy would force the inclusion of phenomenological prescription in the running of the strong coupling. The present formulation only presumes that the leading order dipole picture of DIS is valid, and no other phenomenological modeling is required.

Answers to the significance of the choice of the weight parameter $\lambda$, and the reconstruction method are more elusive.
During the implementation of the noiseless reconstruction we observed that smaller $\lambda$ would correspond to a more accurate recovery of the ground-truth, but once statistical uncertainty is introduced to the data, very small $\lambda$ would begin to over-fit to the noise, and the reconstruction precision would deteriorate and even completely break. 
In badly behaving areas wild fluctuations of the reconstruction or the confidence intervals were seen, and the results would seem to fail to be reasonably physical. This could also manifest as strong tendency to over-fit the data, in which case the ground-truth dipole amplitude was recovered very precisely, but its confidence intervals from the whole $r$-range would explode once the randomly sampled noise is introduced, making the reconstruction highly volatile and unreliable. As a remedy for this, we restricted the weight parameter to the range $\lambda \in [0.0005, 0.01]$.

To construct a more robust reconstruction method for data with experimental uncertainty, we implemented the $\chi^2$-test based error function~\eqref{eq:err-chisq}, which selects the $\lambda$ which most closely fits the data. In practice this is always a larger value for $\lambda$, since the lower values of $\lambda$ will have $\chi^2$ values less than unity. This results in a more stable reconstruction in terms of the point-wise mean of all the randomly sampled reconstructions, that will not over-fit the data, and handles $\lambda$ as a nuisance parameter that is fit to the data.

In the comparison with the other reconstruction and regularization algorithms we compared, we saw that the accuracy and precision of the reconstruction can vary largely based on the used method. We then chose to use the best performing option available, but there very well might be a better performing alternative, or a purpose-built or adapted algorithm could feasibly improve on the present method. One basic improvement would be to adapt the method to be able to work on a logarithmic discretization, which would improve numerical accuracy and could be used to reduce the required resolution of the reconstruction in $r$, and feasibly could also improve the performance of the reconstruction at small $r$.
Finally, one source of reconstruction bias to consider is the regularization method used, i.e. the first order Tikhonov--Phillips algorithm. It is, at least theoretically, biased towards getting the flat peak shape of the dipole amplitude at large $r$, since the regularization penalty grows linearly with the derivative of the reconstruction. And as the reduced cross section data is not highly sensitive to the large $r$ regime of the dipole amplitude, this algorithmic bias can manifest as the reconstructions tending to a constant at large $r$.
As a counter-point in favor of the chosen method, in this closure test where the ground-truth is the fit parametrization using the McLerran--Venugopalan model for the dipole amplitude, we selected the method which best reproduced the ground-truth, i.e. the dipole that flattens off at large $r$. With real data it is probably advisable to compare at least a few methods to gauge the significance of this type of method based bias in the reconstruction.

\section{Conclusions} \label{sec:conclusions}

    In this work, we rewrote the inference of the dipole amplitude from reduced deep inelastic scattering cross section data into a discrete linear inverse problem.
    This novel perspective enables us to consider the extraction of the dipole amplitude from reduced cross section data as a tomographic reconstruction problem without a functional parametrization ansatz for the dipole amplitude that would need to be fit to data.
    This freedom from fit parametrization also enables us for the first time to estimate the confidence intervals of the inference of the dipole amplitude in a more general manner than has been possible previously, and we see that the regimes of high confidence match expectations from the physical theory~\cite{Mantysaari:2018zdd, Casuga:2023dcf, Casuga:2025etc}.
    With this approach, we are pursuing the recovery of the dipole amplitude directly from data in a computed imaging or a (non-)destructive testing sense, with the ultimate aspiration to develop a theory that enables the indirect measurement of the dipole amplitude, and more generally other inferrable quantities as well.

    To build a baseline for this approach, we performed a closure test of the reconstruction process for known parametrizations of the dipole amplitude.
    Specifically, we took the 4- and 5-parameter Bayesian inferences of the dipole amplitude initial condition~\cite{Casuga:2023dcf} as the ''ground-truth`` which we use to simulate reduced cross section data in various fixed Bjorken-$x$ bins using the well-tested numerical implementation used in Refs.~\cite{Ducloue:2017mpb,Beuf:2020dxl,Hanninen:2021byo,Hanninen:2022gje}.
    In the closure test we then perform the numerical reconstruction of the dipole amplitudes from these datasets to demonstrate that our approach and implementation is capable of recovering the fit parametrization dipole amplitudes from the generated reduced cross section data.
    For the numerical reconstruction we use Regtools~\cite{Hansen:1994:regtools,Hansen:2007:regtools} and AIR Tools II~\cite{Hansen:2017airtools}, which are established inverse problems and image reconstruction software packages.

    In the closure test, the reconstruction was quite capable in recovering the ground-truth fit dipole amplitudes especially in the intermediate dipole size $r$ regime of roughly $0.4 - 10 \gev^{-1}$, as seen in Figs.~\ref{fig:plot1-dip-light}, \ref{fig:plot1b-dip-light-loglinear}, and \ref{fig:plot1b-dip-light-ratio}. This is in contrast with the small $r \lesssim 0.3 \gev^{-1}$ and large $r \gtrsim 10 \gev^{-1}$ dipole size regimes, where the reconstruction becomes less accurate, and especially the confidence intervals of the reconstruction sampled with the noisy reconstructions begin to grow notably.
    As discussed in Sec.~\ref{sec:results}, this is expected since in the small-$r$ regime the dipole amplitude vanishes asymptotically and therefore the data becomes unsensitive to minute changes in the amplitude, preventing quantitative reconstruction of the dipole amplitude in the small-$r$ regime. And similarly in the large-$r$ regime the reconstruction loses accuracy, but that is due to the exponential decay of the forward operator at large $r$, again affecting the capability of the reconstruction in that regime.
    The closure test also shows that the reconstruction is able to resolve between the 4- and 5-parameter dipole amplitudes used in the generation of the reduced cross section data, at least in ideal conditions with sufficient or noiseless data. The reconstruction can also recover the normalization $\frac{\sigma_0}{2}$ from the cross section data at least to some level of precision, which is unfortunately affected by the limitation in the large-$r$ regime discussed above.
    Lastly, we verified that the reconstruction performs equally well when the contribution from the charm quark is included, since that does not change the nature of the underlying mathematical inverse problem, and only produces a minor correction to the forward operator.

    This \edit{proof-of-principle} closure test with precisely known dipole amplitudes will help us to understand the reconstruction results in more detail, once we apply this framework to HERA DIS data.
    The reconstruction of the dipole amplitude from real data seems to hold great potential for \edit{novel small-$x$} phenomenology\edit{, such as an in-detail comparison between the reconstructed dipole amplitude, and theoretical predictions for the Bjorken-$x$ evolution of the dipole amplitude as described by evolution equations such as the Balitsky--Kovchegov~\cite{Balitsky:1995ub,Kovchegov:1999ua,Kovchegov:1999yj} or JIMWLK~\cite{JalilianMarian:1996xn, JalilianMarian:1997jx, Jalilian-Marian:1997ubg, JalilianMarian:1997gr, Iancu:2001md, Iancu:2000hn, Ferreiro:2001qy, Iancu:2001ad} equation. This would be a novel opportunity to examine the more computationally expensive evolution equations, since their predictions could be compared directly against the dipole amplitude reconstructed from the cross section data, instead of having to use them in a conventional fit procedure.}
    Or alternatively, if the reconstructions are evolved \textit{backward} towards larger Bjorken-$x$ \edit{with the BK equation}, does the \edit{resulting} ''initial condition`` functional shape agree with the theoretically motivated models, such as McLerran--Venugopalan or IP-sat parametrizations of the initial condition?
    It will be interesting to see, whether we are able to reliably reconstruct the normalization $\sigma_0(\xbj)$ from inclusive DIS HERA data. That would give a view free from parametrization-bias into the transverse area of the proton as the function of Bjorken-$x$, since the reconstruction is performed independently at each $x$.
    The application of this reconstruction framework to inclusive HERA data is on-going and out of the scope of this proof-of-principle work.
    \edit{While the available HERA data could prove to be a challenging application for these methods, the next-generation data from the Electron--Ion Collider should be of sufficient fidelity and quantity for this approach.}

    The novel approach employed in this work is enabled by the mathematical inverse problems paradigm, where the underlying mathematical nature of the connection between the inferred quantity and the measured data is leveraged to solve the inference problem in a more general sense, as is done in indirect measurement and imaging. The key step was to recognize the inclusive DIS in the dipole picture as a linear integral transformation inverse problem, to which applicable standard methods are widely available from numerous applications~\cite{Hansen:1994:regtools,Hansen:2007:regtools,Hansen:2017airtools,Hansen:2021ct}.
    \edit{In a sense, this approach is a generalization of the perspective taken in Ref.~\cite{Munier:2001nr} where the analytic invertibility of the Fourier transform is leveraged towards analogous model-free inference of the impact parameter dependence of the dipole amplitude, whereas in this work we employ methods applicable to a more general class of integral transformation problems to infer the $r$-dependence.}
    
    The inverse problems paradigm is by no means limited to inference problems such as the one underlying the dipole amplitude inference problem, and could be a fruitful approach in high-energy physics. For some applications a similar distillation of the explicit inverse problem as was done in this work could feasibly be enough to open new opportunities, and for others novel mathematical theory could be required for the development of numerical algorithms, for example.
    Theory understanding of the underlying mathematical inverse problem can enable novel approaches, such as developed in this work, and on the other hand give insight into what would be an efficient measurement in the experiment. For example, the current implementation of the dipole amplitude reconstruction would benefit from having a lot of data points in $Q^2$ at fixed $\xbj$ and $\sqrt{s}$. These types of observations will be inference problem and method specific, however.

    More fundamentally, this work leads us to questions about what is measurable as we begin formulating this inverse problems approach into a theory of indirect measurement that can be applied to inference problems in high-energy physics.
    For example, consider the reconstruction performed in computational tomography, where the 3-dimensional structure of a subject is recovered from a dataset of 2-dimensional projections of the structure taken by x-ray imaging. The latter seems quite clearly to be a measurement, and so would seem the reconstructed 3-dimensional structure as well, as it can be used for quantified observation of the interior.
    We bring up this example, because the mathematical inverse problem that is solved in the reconstruction of the 3-dimensional structure happens to be highly analogous to the dipole amplitude inverse problem considered in this work, and in fact the very methods employed in this work can be applied to computational tomography.
    Now, consider the analogous steps of the dipole amplitude inverse problem: the measurement of the reduced cross section in inclusive DIS, and the reconstruction of the dipole amplitude based on the same type of discrete linear inverse problem and numerical implementation. Does this mathematical correspondence between the computed tomography and dipole amplitude inverse problems imply, that the reconstruction of the dipole amplitude from reduced cross section data would amount to a measurement of the dipole amplitude?
    This could perhaps be akin to a measurement of a probability distribution of a physical process, which is a quantifiable task, but the probability distribution itself is not a physical observable?
    These questions come into focus in our next article, in which we perform the first inverse problem reconstruction of the dipole amplitude from HERA deep inelastic scattering data.

    With this inverse problems approach to QCD phenomenology, we are building a path towards a new generation of high-energy phenomenology, where inferrable quantities such as the dipole amplitude are reconstructed---or even measured indirectly---from experimental data. \edit{With the future Electron--Ion Collider producing significantly higher fidelity data than is available currently, now is the right time to develop more sophisticated methods of inference by leveraging established methods of inverse problems.} This would open a new opportunity for theoretical and phenomenological work in the description of the reconstructed quantity from more fundamental degrees of freedom in QCD.

\vspace{3mm}
\noindent
\textbf{Code availability}
The source code \cite{oma4:invdip-github} for the discretization of the forward operator, and for the reconstruction implementation, is available at \url{https://github.com/hhannine/inversedipole/tree/arXiv-2509.05005}. 

\vspace{3mm}
\noindent
\textbf{Acknowledgments} 
We express our gratitude to J. Ilmavirta, H. Mäntysaari, B.M. Afkham, M. Kuha, and K. Kansanen for fruitful discussions and comments.
H.H. is supported by the Research Council of Finland (Flagship of Advanced Mathematics for Sensing Imaging and Modelling grant 359208; Centre of Excellence of Inverse Modelling and Imaging grant 35309), and the Vilho, Yrjö and Kalle Väisälä Foundation.
A.K. was supported by the Geo-Mathematical Imaging Group at Rice University.
H.S. is supported by the Research Council of Finland (Flagship of Advanced Mathematics for Sensing Imaging and Modelling grant 359208).

\bibliographystyle{JHEP-2modlong.bst}
\bibliography{refs}

\end{document}